# Two hypotheses about natural intelligence

Claudio Parmeggiani

*CEAP – Via Balzaretti, 24 – 20133 Milano (Italy)*

*E-mail: clparm@infinito.it*

**Abstract**. We assume that the natural intelligence (human, particularly) is equivalent to a large inferring structure, which took shape in the last 400/500 million years. Then two hypotheses, about this structure and its development, are put forward for consideration. The first one concerns the transmission, from one generation to another, of the structure: we propose that this passage is done by direct transfer, mother to children, during pregnancy ("maternal download"). The second hypothesis regards the structure evolution: now the acquired improvements can be transferred to the descendants, so it is possible to envisage a governed evolutionary process ("evolution by improvements").

## Introduction

Following [*1*] and, to some extent [*3*], we assume that the natural intelligence is equivalent to an inferring structure (an articulate and organized object able to infer and compute), presently very large and complex, at least for the mammals. Because all such structures are in a sense equivalent [*1, 2*], this assumption is not really objectionable; the only uncertain point is the structure size. This size will be evaluated employing, as unit of measure, the usual programming instructions, but every other method could be considered appropriate. A size, for today's humans, of $10^9 \div 10^{10}$ instructions seems reasonable and, in any case, compatible with the physical characteristics of the brain. Moreover, the whole structure (or great part of it) must have taken shape, in some manner, by some evolutionary process, in the last 400/500 million years. Now, we are asking ourselves: (A) how the inferring structure passes from one generation to another; (B) in which manner it has reached the current size and complexity, at least for some species. In the first and second section we propose an answer, based upon two, strictly connected, hypotheses: (A) the generation-to-generation transmission is done by direct transfer ("maternal download") from the mother to the children, during pregnancy; (B) every female of the specie bring some improvements and, because of (A), contribute to the structure global evolution ("evolution by improvements"). In the third and forth section we examine some implications of the hypotheses, on the structure organization, growth and uniformity (at population level).

## 1. The Maternal Download hypothesis

We propose that the inferring structure (which represents natural intelligence) is memorized on the brain (perhaps as a long-term memory); then during pregnancy it is duplicated and transferred to the children; the genetic code provides, eventually, the rules for the transfer and some functional parameters. This is the maternal download hypothesis. Without this hypothesis, we can explain the generation-to-generation transmission only by the following alternatives:

(GEN) the whole structure is genetically coded and, therefore, transferred from (both) the parents to the children, at conception.

(ENV) the structure, as a matter of fact, does not pass between generations, but it is built up after the birth, with the help of environmental interactions. From the genetic code we obtain only the build up general principles; however these "initial data" cannot determine univocally the final result: if so it was, we would have only a different version of (GEN).

Both (GEN) and (ENV) present some problems:

(GEN1) the structure size and complexity are such that a genetic coding is (physically) impossible, also in some compressed manner; on the contrary, the structure memorization at brain level seems easily feasible. Obviously it is possible that simpler structures (typically with only "reactive" functions) are entirely genetically coded.

(GEN2) aside from the transfer method, the instructions have to be, first, duplicated, very accurately. The genetic coding assumption requires a quite fast duplication; the download hypothesis allows a much slower duplication, of about a few million times.

(ENV1) in consequence of structure size and complexity, a reliable after birth development, without evolution, seems all but impossible: we have at our disposal only few years, not the million years of the evolutionary process implied by the download hypothesis.

(ENV2) the after birth assumption do not account for the fact that all individuals, at least within a single population, have almost the same inferring structure (population uniformity). Without this "oneness of the mind" the population survival appears very precarious; in any case, we notice, in nature, a high degree of uniformity, also for all the mammals.

Apparently the download hypothesis explains everything, a part the populations' uniformity. It is necessary, in this case, a supplementary assumption: we must admit that today's large populations come from few individuals and that the growth process has been relatively fast. The argument will be discussed in the forth section, Population Uniformity and Divergence.

## 2. The Evolution by Improvements hypothesis

We now propose that the inferring structure (which represents intelligence and is memorized on the brain) is modified (enlarged and improved, mainly) along the whole individual's life; in this (somewhat controlled) process environmental interactions play a considerable role, of stimulus and guidance; the whole structure is then copied and transferred, modifications included, during pregnancy. This is the "evolution by improvements" hypothesis. Without this hypothesis, we can explain the structure growth, and its current size and complexity, only by:

(MUT) the whole inferring structure (or the guiding rules for the after birth development) is genetically coded; random mutations and recombination processes improve or worsen it; the environmental selection preserves and amplifies the improvements.

The (MUT) assumption presents some critical points, in connection both with (GEN) and (ENV):

(MUT1) random mutations and recombination processes seem unsuited to produce such a kind of structure: the result should be very often not operating, making the selection process exceedingly slow. In any case we do not observe, in nature, such a high rate of negative results. On the contrary, the life-long structure improvements are, in some degree, guided and controlled; besides, in consequence of the download, there is not any unforeseeable recombination.

(MUT2) if the structure is (mostly) developed after individual's birth and the results are irremediably lost at individual's death, every evolutionary process is frozen. The mere evolution of the "guiding rules" cannot significantly contribute.

Generally, if the download assumption is, at least partially, right the evolution by improvements hypothesis seems unavoidable (and unobjectionable). An interesting collateral aspect of the hypothesis is the possibility to generate improvements that are not immediately useful. Finally, we ought to observe that mutations (acting upon physical structures) and inferring improvements are quite interdependent. From one side a mutation induces inferring improvements such that the resulting structure could make a better use of the new or modified physical trait. From the other side, an inferring improvement favours, by environmental selection, the mutations coherent or consistent with it.

## 3. Structure Organization and Growth

The whole structure can be thought as composed of many substructures, altogether independent, each one with the size, more or less, of a few million instructions; the evolution of substructures has lasted, say, 500 million years. Moreover, we assume that the substructures creation rate, $\sigma$, is constant in time and that every substructure evolves independently. If $\rho(x)$ is the growth rate (instructions per time unit) of the substructure $x$, the total production rate (of instructions) is, at time $T$:

$$V(T) = \sum_{x \in X(T)} \rho(x) = \sigma . T . \overline{\rho} \tag{1}$$

Here $X(T)$ is the set (with $\sigma.T$ elements) of all substructures present at time $T$; $\overline{\rho}$ is the arithmetic mean of growth rates, for all substructures. When $T = 0$ (at the beginning of the development, when there are not structures and instructions) we assume that also $V = 0$. The total amount of instructions, at time $T$, is consequently:

$$S(T) = \frac{1}{2} \sigma . \overline{\rho} . T^2 \tag{2}$$

Posing $\sigma = 10^{-6}$ substructures per year and $\overline{\rho} = 10^{-2}$ instructions per substructure and per year, we have (if $T = 5.10^8$ years) $S(T) \approx 10^9$ instructions and $V(T) = 5$ instructions per year. Both values seem reasonable, perhaps underrated, for the humans. The amount of existing substructures is then 500.

Distinct substructures could have, at a given time, quite different growth rates: some are steady and therefore devoted to well established tasks; others are fully evolving, continuously modified and assigned to tasks "in course of definition". A parallel is suggested [3] between steady structures and unconscious components of the mind; and between dynamic structures and consciousness ("consciousness is a phenomenon in the zone of evolution"). But, try to embody this idea, seems not so easy (and perhaps premature).

Moreover, the production rate of instructions could change during individual's life. In such a case the (average) generation and life lengths would affect the total production rate: if the individual production rate increase/decrease during the life, the populations with a large generation length (with respect to life length) are favoured/disfavoured. Suppose, for example, that the individual production rate (of instructions) is proportional to $t^\alpha$; $t$ is the time, measured from the individual's birth and $\alpha \geq 0$. If $s(t)$ is the amount of instructions produced at time $t$, if $t_g$ and $t_h$ are the average generation and life lengths and $\xi = t_g/t_h$, we have:

$$\frac{s(t_g)}{s(t_h)} = \left(\frac{t_g}{t_h}\right)^{\alpha+1} = \xi^{\alpha+1} \qquad (3)$$

After $T/t_g$ generations, the total amount of instructions produced, within population $A$, is:

$$S_A(T) = \frac{s(t_h)}{t_h} \cdot \left(\frac{t_g}{t_h}\right)^\alpha \cdot T = v(T) \cdot \xi^\alpha \cdot T \qquad (4)$$

where $v(T) = s(t_h)/t_h$ is an average production rate of instructions, during the individual's life; $v$ generally depends on $T$. For two populations, $A$ and $B$, with different $\xi$ (but the same $v$) we have $S_A/S_B = (\xi_A/\xi_B)^\alpha$; this relation implies (if $\alpha > 0$) that the final result (the "average" population's intelligence) is very dissimilar, for $A$ and $B$.

We can also imagine that different kinds of instructions have distinct production rates. Let assume, say, that the "speculative" instructions have a rate proportional to $t^\beta$; $t$ is always the time, measured from the individual's birth and $\beta > \alpha$. If $S_A^*(T)$ is the total amount of speculative instructions, produced within the population $A$ after $T/t_g$ generations, then:

$$R_A(T) = \frac{S_A^*(T)}{S_A(T)} = \frac{v^*(T)}{v(T)} \cdot \xi^{\beta-\alpha} \tag{5}$$

Obviously $v^*(T)/v(T) = s^*(t_h)/s(t_h) < 1$ so also $R_A(T) < 1$. For two populations, $A$ and $B$, with different $\xi$ (but the same $v$, $v^*$) we have $R_A/R_B = (\xi_A/\xi_B)^{\beta-\alpha}$, with evident consequences on the speculative capabilities of $A$ and $B$.

Till now we have tacitly admitted that all modifications are only improvements, but it is also possible that some (few) of them are "worsening". Let be $P_k(t)$ the probability of $k \geq 0$ worsening at time $t$ (measured from the individual's birth); $P_k$ is, reasonably, a Poisson process, so:

$$P_k(t) = \frac{(F(t))^k}{k!} e^{-F(t)} \tag{6}$$

Here $F$ is a positive, increasing function and $F(0) = 0$. The probability of some $(\geq 1)$ worsening at time $t$ is then $1 - P_0(t) \approx F(t)$, for small $t$. Finally, the probability of a number of worsening $\geq k$, at time $t$, is:

$$Q_k(t) = 1 - P_0(t) - P_1(t) - \ldots - P_{k-1}(t) \tag{7}$$

We can assume that, after too many worsening, some degenerative processes and impairments take place. Choosing suitable $k$ and $F$, we can obtain a $Q_k$ with some resemblance to prevalence curves of senile degenerative illness; for example, we can pose: $k = 8 \div 10$; $F(t) = (p.t)^{1+a}/(1+a)$ and $a = .85 \div .70$, $p = (4 \div 5).10^{-2}$ year$^{-1}$. More, in this case, for $t \approx 100$ (or larger) and $k \geq 10$, $Q_k(t)$ is very near to 1.

## 4. Populations Uniformity and Divergence

A population is said "uniform" if all individuals have inferring structures very similar; by ours hypotheses this is only possible if the majority of individuals have a unique (female) ancestor, not to far in time. Not to far because, if not, the growth process described below would generate dissimilarities (divergences) which could destroy the uniformity. If $m_1(a)$ is the mother (father) of the female (male) $a$, we pose $m_{n+1}(a) = m_1(m_n(a))$ for $n \geq 1$; $m_n(a)$ has the same sex of $a$ and is her (his) $n - ancestor$. We now consider, for a population $A$, the probability $P_n(A)$ that two whatever individuals $a$ and $b$, $a \neq b$, of the same sex, have the same $n - ancestor$ (that is $m_k(a) = m_k(b)$ for some $k \leq n$, $n \geq 1$):

$$P_n(A) = 1 - \prod_{0 \leq k \leq n-1}(1 - P_1(A_k)) \tag{8}$$

where $A_k$ is the population $k \geq 0$ generations before and $A = A_0$.

$P_1(A)$ is the probability that two whatever (but different) females (males) of $A$ have the same mother (father). It depends on $N_1(A), 2.N_2(A), \ldots$ which are the numbers of females (males) of $A$ with $0, 1, \ldots$ sisters (brothers):

$$P_1(A) = \frac{\sum_{k \geq 2} k(k-1).N_k(A)}{N(A).(N(A)-1)} = \frac{M(A)}{N(A)} \tag{9}$$

Here $N(A) = N_1(A) + 2.N_2(A) + \ldots$ is total number of females (males) of $A$. The above $M(A) \approx \sum_{k \geq 2}(k^2 - k)(N_k(A)/N(A))$, the "population index", is, generally, quite near to 1, but in some cases can differ from it sensibly. This is the case, for example, if only few males (females) are allowed to have children, or if every male (female) has almost exactly one son (daughter). We are interested only to $N(A)$ and $n$ quite large so:

$$P_n(A) \approx 1 - \exp(-\sum_{0 \leq k \leq n-1} P_1(A_k)) = 1 - \exp(-n\frac{\overline{M(n)}}{\overline{N(n)}}) \tag{10}$$

Here $\overline{N}(n) = \sum_{0 \le k \le n-1} M(A_k) \Big/ \sum_{0 \le k \le n-1} (M(A_k)/N(A_k))$ is a weighted harmonic mean (over $n$ generations) of female (or male) population sizes and $\overline{M}(n)$ is the arithmetic mean of population indices. The above sums can be, generally, replaced by integrals. If $n \approx \overline{N}(n)/\overline{M}(n)$ (or greater) $P_n(A) \approx 1$; for example with $n = 5.(\overline{N}(n)/\overline{M}(n))$ we have $P_n(A) = 0.993$.

But $n$ cannot be too large: if so it were the growth process would bring too many dissimilarities. We can estimate this $n$-generations "divergence" by the ratio:

$$R_n = \frac{S_n}{S_0} = \left(1 - \frac{n}{n_0}\right)^{\gamma} \approx 1 - \gamma \cdot \frac{n}{n_0} \qquad (11)$$

where $S_0$ and $S_n$ are the amounts of inferring instructions, current and $n$ generations before; $n_0$ is the number of generations from the beginning of the structure development; $n_0 \approx 3.10^7$ and $\gamma = 2$, by our preceding assumptions. If, for example, $n = 5 \cdot 10^4$, we have $R_n = 0.997$. Finally, we can define a "uniformity measure", $U(A)$, for a population $A$, as the maximum of some average of the female $P_n(A)$ and of $R_n(A)$; for example:

$$U(A) = \max_{1 \le n \le n_0} \frac{P_n(A) + R_n(A)}{2} \qquad (12)$$

We can now draw the following conclusions:

(CONST) If population size and index, $N$ and $M$, are almost constant, we have uniformity for a small ($10^{-3} \div 10^{-4}$) $K = \overline{N}/(\overline{M} \, n_0)$. In this case $n_{max} = -n_0 K . \log(2K)$ and $U = 1 - K(1 - \log(2K))$. Small populations are always uniform, but large ones (for example today's humans) are fated to diverge.

(LIN) For a (linearly) shrinking population, $N(A_k) = N_0 + k/\lambda$. If $M$ is constant (but it is easy to consider also the linear case), we have $P_n \approx 1 - (1 + n/(\lambda N_0))^{-\lambda M}$, for large $n$ and $N_0$.

This probability is extremely sensible to the value of $\lambda M$. For $\lambda M = 1$ the results are quite simple: $n_{max} = n_0 (\sqrt{K/2} - K)$, $U = 1 - \sqrt{2K} + K$ and $K = N_0 /(M\, n_0)$ must be small (to have uniformity).

(EXP) For an (exponentially) expanding population, $N(A_k) = N_0\, e^{-k/\lambda}$. If $M$ is constant (or also exponential, $M(A_k) = M_0\, e^{k/\mu}$), we have, posing $1/\lambda^* = 1/\lambda + 1/\mu$,

$P_n \approx 1 - \exp(-(\lambda^* M_0 / N_0)(e^{n/\lambda^*} - 1))$, for large $n$ and $N_0$ (but $N_0 >> e^{n/\lambda^*}$). A possible choice of parameters, adapted to humans recent evolution, is the following: $M_0 \approx 1$, $N_{current} = N_0 = 10^9$, $N_{initial} = 500$, $\lambda \approx 700$, $\mu \approx 10^4$. Consequently, $P_n = 0.98$ and the number of generations from the "beginning" to now is: $\lambda \log(N_{current} / N_{initial}) \approx 10^4$. We obtain a perhaps better model if, initially (*n* generations before), population size and index are kept constant.